\documentclass{article}

\usepackage{textcomp}
\usepackage{hyperref}
\usepackage{lipsum}

\usepackage{spconf,amsmath,graphicx}

\usepackage{booktabs}       
\usepackage{ amssymb }
\usepackage{amsmath}
\DeclareMathOperator*{\argmax}{argmax} 
\usepackage{xcolor}
\usepackage{multirow}

\usepackage[absolute,showboxes]{textpos}

\TPMargin{5pt}

\newcommand{\copyrightstatement}{
    \begin{textblock}{0.84}(0.08,0.93)    
         \noindent
         \footnotesize
         \copyright 2020 IEEE. Personal use of this material is permitted.
  Permission from IEEE must be obtained for all other uses, in any current or future 
  media, including reprinting/republishing this material for advertising or promotional 
  purposes, creating new collective works, for resale or redistribution to servers or 
  lists, or reuse of any copyrighted component of this work in other works. 
  DOI: \href{https://doi.org/10.1109/ICASSP40776.2020.9053662}{10.1109/ICASSP40776.2020.9053662}
    \end{textblock}
}

\title{LAI-Net: Local-Ancestry Inference with Neural Networks}

\name{Daniel Mas Montserrat$^{\star}$, Carlos Bustamante$^{\dagger}$, Alexander Ioannidis$^{\dagger}$}
\address{$^{\star}$Purdue University \quad  $^{\dagger}$Stanford University}

\begin{document}

%


\maketitle
\copyrightstatement

\begin{abstract}
Local-ancestry inference (LAI), also referred to as ancestry deconvolution, provides high-resolution ancestry estimation along the human genome. In both research and industry, LAI is emerging as a critical step in DNA sequence analysis with applications extending from polygenic risk scores (used to predict traits in embryos and disease risk in adults) to genome-wide association studies, and from pharmacogenomics to inference of human population history.
While many LAI methods have been developed, advances in computing hardware (GPUs) combined with machine learning techniques, such as neural networks, are enabling the development of new methods that are fast, robust and easily shared and stored. In this paper we develop the first neural network based LAI method, named LAI-Net, providing competitive accuracy with state-of-the-art methods and robustness to missing or noisy data, while having a small number of layers.
\end{abstract}
\begin{keywords} 
Local-Ancestry Inference, Genomics, Genetics, Neural Networks, Deep Learning
\end{keywords}

\section{Introduction}
\label{sec:intro}
\begin{figure}
    \centering
        \label{fig:lai}
        \includegraphics[width=0.99\linewidth]{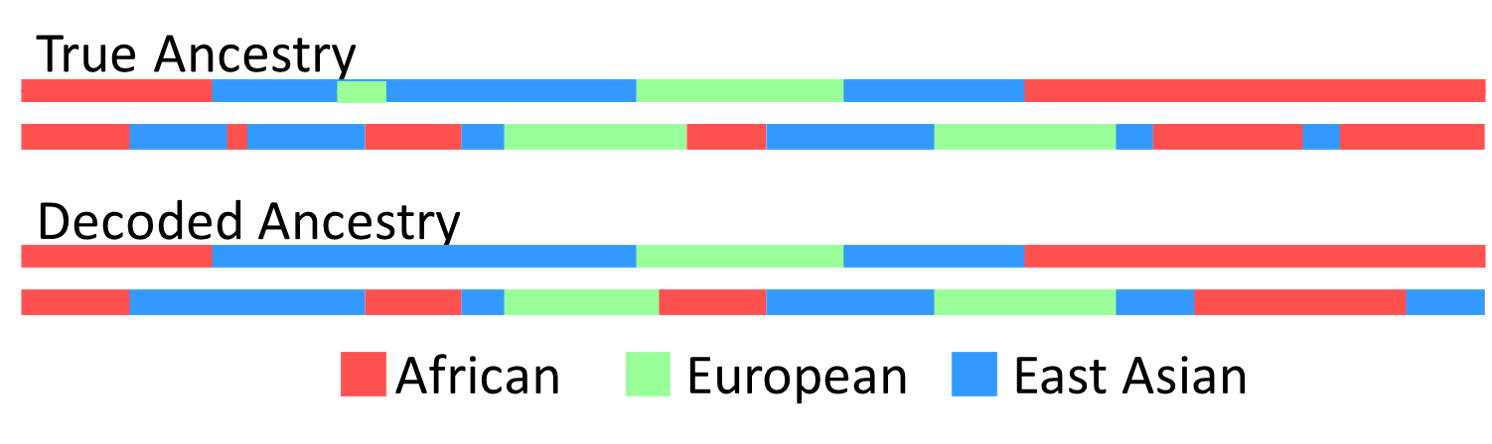}
    \caption{Illustration of local-ancestry inference problem. An admixed
pair of chromosomes are shown with the true ancestry (top) and the decoded
ancestry (bottom). The admixed diploid individual has three ancestral population sources: African, European and East Asian.}
\end{figure}

Although most positions in the human DNA sequence (genome) do not vary between individuals, about two percent ($\sim$5 million positions) do; these are referred to as single nucleotide polymorphisms (SNPs) and can be encoded as binary variables with zero denoting the common variant and one denoting the minority variant. Modern human populations--originating from different continents and different subcontinental regions--exhibit discernible differences in the frequencies of SNP variants at each position, and in the correlations between these variants at different nearby positions, due to genetic drift and differing demographic histories (bottlenecks, expansions and admixture) over the past fifty thousand years \cite{Li:2008ena, DeGiorgio:2009cs}. Local-ancestry inference (illustrated in Figure \ref{fig:lai}) uses the pattern of variation observed at such sites along an individual's genome to estimate the ancestral origin of each segment of an individual's DNA. Because DNA is inherited as an intact sequence with only rare, random swaps in ancestry (between the two parental DNA sequences) at each generation, ancestral SNPs form contiguous segments allowing for powerful ancestry inference based on patterns of contiguous SNP variants.

Recent advances in machine learning, together with the growth in number and density of genome-wide training sets, are allowing for the identification and classification of ancestry-specific patterns of sequences of variants along the DNA strand with increasing accuracy \cite{Price:2009bga}. Indeed, ancestry assignment is now often made for each segment of an individual's DNA at milliMorgan resolution \cite{maples2013rfmix}. This type of high-resolution local-ancestry inference is becoming an important part of medical association studies \cite{martin2017unexpectedly}, human demographic inference \cite{MorenoEstrada:2014jr}, and even consumer products \cite{Durand:2015jx}. 

The ability to accurately infer the ancestry along the genome in high-resolution is important to disentangle the role of genetics and environment for complex traits including illness predisposition, since populations with a common ancestry share complex physical and medical traits. For example, Puerto Ricans living in the United States have the highest mortality of asthma and Mexicans have the lowest \cite{torgerson2012case}. Elucidating the genetic associations within populations for biomedical traits (like height, blood pressure, cholesterol levels, and predisposition to certain illness) can inform the development of treatments, and allow for the building of predictors of disease risk, known as polygenic risk scores. However, because the correlations between neighboring genetic variants are ancestry dependent, applying these risk scores to an individual's genome requires knowledge of the individual's ancestry at each site along the genome. With the increasing diversity of admixed modern cosmopolitan populations, such ancestry-specific analysis along the genome is becoming an increasingly complex and important computational problem.

Many different LAI methods have been developed. Methods such as SABER \cite{Tang2006}, HAPAA \cite{Sundquist2008} and HAPMIX \cite{Price:2009bga} model local-ancestry correlations with Hidden Markov Models (HMMs). The LAMP algorithm \cite{sankararaman2008estimating} utilizes probability maximization within a sliding window providing better and faster performance than several HMMs-based models, even in recently admixed populations. RFMix \cite{maples2013rfmix} is a discriminative model that uses conditional random fields (CRF) based on random forests within windowed sections of the genome. RFMix provides state-of-the-art results and has shown to be both faster and more accurate than LAMP and previous HMMs-based methods.

Analyzing human genomic sequences can be challenging as the data is high-dimensional, while available annotated datasets typically contain only a small number of samples. Therefore, methods that are capable of processing high-dimensional information in an efficient manner are preferred. Additionally, many datasets containing human genomic sequences are proprietary, protected by privacy restrictions, or are otherwise not accessible to the public. Models that can be easily shared once trained can be useful in such scenarios. While the datasets with their de-identifiable genome-wide sequences remain securely private, models trained on them could be made publicly available. 

In recent years, deep learning has proved useful in solving computer vision and natural language processing problems, \cite{nature} and it is becoming widespread in the medical field. From analyzing MRI scans \cite{lundervold2019overview} and detecting tumors within images \cite{Gar2016} to finding disease predisposition in the human genome \cite{li2018heterogeneity}, neural networks have provided useful and effective solutions. Several deep learning methods have recently been presented in the field of genomics \cite{vaegan2019, eraslan2019deep}.

In this work we present a neural network named LAI-Net and its lightweight version named Small LAI-Net. Both networks achieve state-of-the-art results on admixed individuals simulated from real human sequences. Additionally, experimental results show that the networks are robust to missing data and phasing errors.


\section{Admixture Simulation Dataset}




In this work we use full genome sequences obtained from human research participants through the 1000 genomes project \cite{10002015global}. We select a total of 1668 single-population individuals from East Asia (EAS), African (AFR) and European (EUR) ancestry. The East Asian group is composed of the following individuals: 103 Han Chinese in Beijing, China (CHB), 104 Japanese in Tokyo, Japan (JPT), 105 Southern Han Chinese (CHS), 93 Chinese Dai in Xishuangbanna, China (CDX) and 99 Kinh in Ho Chi Minh City, Vietnam (KHV). The African group is composed of the following individuals: 108 Yoruba in Ibadan, Nigeria (YRI), 99 Luhya in Webuye, Kenya (LWK), 113 Gambian in Western Divisions in the Gambia (GWD), 85 Mende in Sierra Leone (MSL), 99 Esan in Nigeria (ESN), 61 Americans of African Ancestry in Southwest USA (ASW) and 96 African Caribbeans in Barbados (ACB). Finally, the European group is composed of the following sub-populations: 99 Utah Residents (CEPH) with Northern and Western European Ancestry (CEU), 107 Toscani in Italia (TSI), 99 Finnish in Finland (FIN), 91 British in England and Scotland (GBR) and 107 Iberian Population in Spain (IBS).


Using the full genomes of these individuals we simulated admixed descendants using Wright-Fisher forward simulation over a series of generations. In particular, from the 1668 single-population individuals, 1328 were selected to generate 600 admixed individuals for training, 170 were used to generate 400 admixed individuals for validation and the remaining 170 were used to generate 400 admixed individuals for testing. 
The validation and testing set was generated using 10 individuals for each of the 17 different ancestries. The 600 admixed individuals of the training set were composed by groups of 100 individuals generated after 2, 4, 16, 32 and 64 generations. The 400 admixed individuals of the validation and testing set were generated with 6, 12, 24 and 48 generations each. (With increasing numbers of generations following initial admixture, descendants have increasing numbers of ancestry switches along the genome, leading to a more challenging inference.) This simulation scheme allowed for training and testing the network over a wide range of generations; yielding a method that is robust to populations and individuals having different admixture histories.

\section{Neural Network Architecture}
The proposed network, LAI-Net, is composed of two sub-network: a classification network and a smoothing layer. The network is trained to infer the ancestry of phased diploid sequences. The first layer provides an initial ancestry estimate, $\hat{y}_1$, within windowed regions of the chromosome sequence. The second layer smooths the estimates over multiple windows providing the final estimate $\hat{y}_2$. Figure 2 presents the network architecture.

The first sub-network consists of a set of classifiers within non-overlapping windows of a fixed size of 500 SNPs. The input of the network consists of the base-pairs of the SNPs encoded as -1 (common variant) and 1 (minority variant). Each classifier is composed of a linear layer of size 500 $\times$ 30 followed by a ReLU activation and batch normalization. A linear layer of size 30 $\times$ $N_A$, combined with a softmax function, that maps the hidden layer to the probabilities for assignment to each of the possible ancestries (in this case $N_A=3$, African, European and East Asian). The first sub-network is used twice, one for each sequence of the diploid individual. The second sub-network, a smoothing layer, consists of a two-dimensional convolution layer that takes as inputs the concatenated probabilities of the first layer and outputs the classification estimation within each window. The convolution layer has a kernel size of 75 $\times$ 2 and $N_A$ input and output channels. Therefore, the ancestry of each window is inferred by weighting the 75 initial neighboring estimates for both maternal and paternal sequences. The convolution is performed with the proper reflection padding in order to maintain the same input and output size of the layer. By using a convolutional layer we obtain invariance of the order in which both sequences are presented (i.e. the output of the network is the same, up to a permutation, independently if the maternal or paternal sequence is presented first or last).





The network is trained with two cross-entropy loss functions:
   $\mathcal{L}(y,\hat{y}) = \lambda_1 \mathcal{L}_{CE}(y,\hat{y}_1) + \lambda_2 \mathcal{L}_{CE}(y,\hat{y}_2)$
The first loss function, $\mathcal{L}_{CE}(y,\hat{y}_1)$, compares the estimate of the first sub-network with the true ancestries and updates the weights of the first sub-network. The second term, $\mathcal{L}_{CE}(y,\hat{y}_2)$, compares the estimate of the last layer with the true ancestries and updates the weights of the overall network. When $\lambda_1 > 0$, the output of the first layer, $\hat{y}_1$, represents the probabilities estimated by the classifiers, otherwise the output of the classifiers can be interpreted as a hidden layer. In this work we use $\lambda_1 = \lambda_2 = \frac{1}{2}$.

Dropout regularization is applied to the input data. This models missing input SNPs and provides robustness to missing data, which is a common occurrence when using current commercial genotyping arrays. Experimental results, presented in section \ref{sec:experimental-results}, suggest that even with half of the SNP sites removed (treated as missing), the network is able to accurately estimate ancestry, suggesting that models trained on one genotyping array could even be applied to another genotyping array. (Different commercial genotyping arrays sequence different sets of SNPs with intersections as low as 50\%.)

\begin{figure}
 \label{fig:lainet}
    \centering
    \hspace{-0.5cm}
        \includegraphics[width=0.99\linewidth]{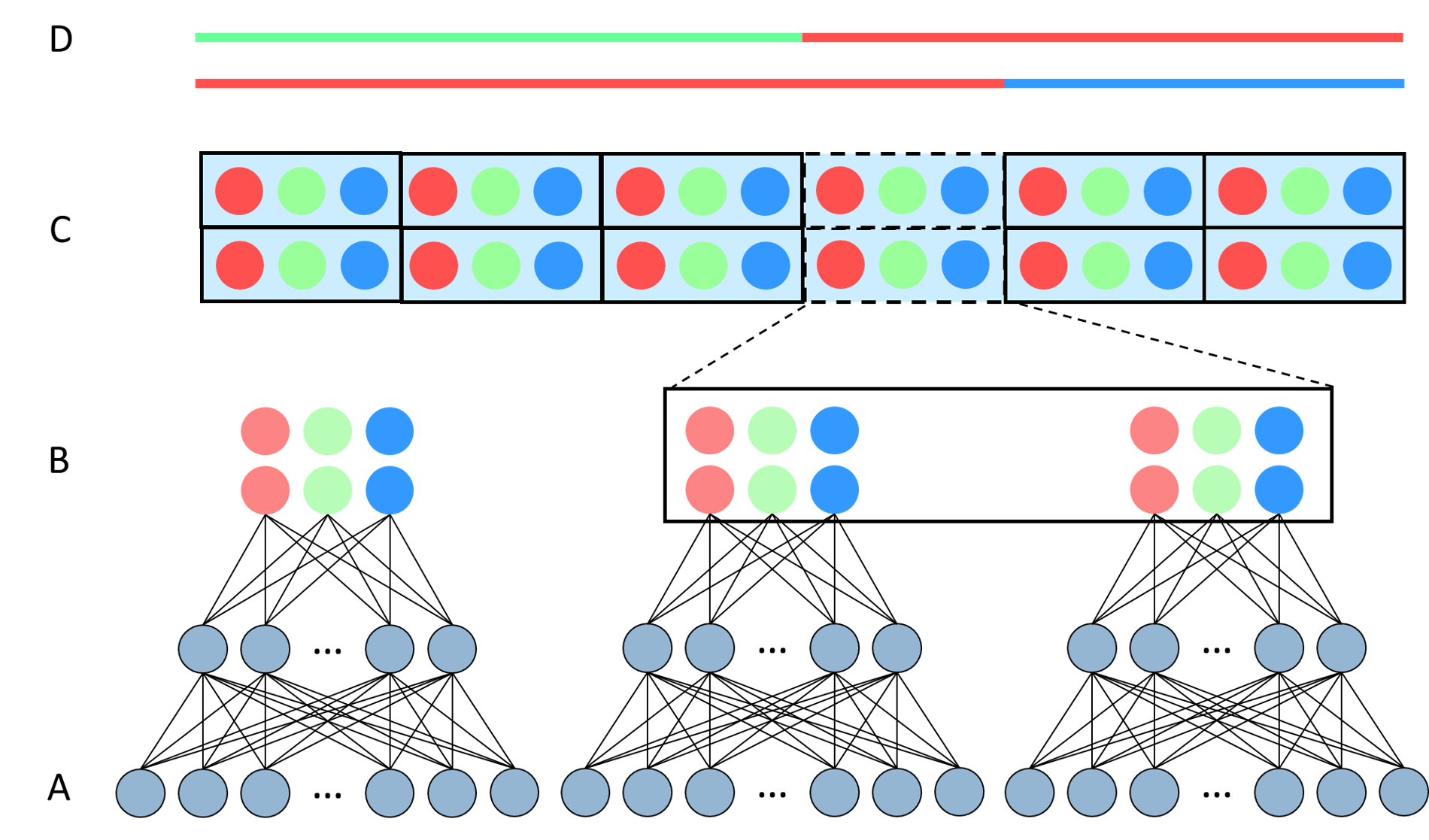}
    \caption{LAI-Net architecture. \textbf{A:} Input SNPs, \textbf{B:} Output of first layer, $\hat{y}_1$, for maternal and paternal sequence, \textbf{C:} Output of smoothing layer, $\hat{y}_2$, \textbf{D:} Inferred ancestry at each window, $\argmax \hat{y}_2$, for maternal and paternal sequence. Each color represents a different ancestry (AFR, EUR, EAS).
    }
     \vspace{-0.1cm}
\end{figure}

While methods such as RFMix require the user to specify the number of generations since admixture, LAI-Net can handle populations of variable, or unknown, generations since admixture. Generation agnosticism is obtained by training the network with data simulated over a wide range of generations. However, even if only one generation is used for training, experimental results suggest that the network is still able to infer ancestry from other generations with only a small decrease of accuracy.

\subsection{Small LAI-Net}
We present a lightweight version of the network that we name Small LAI-Net. This network follows the same scheme as LAI-Net, but with the hidden layer removed. Thus, the network is composed only of two layers: a set of linear classifiers with size 500 $\times$ $N_A$ and a smoothing layer with a kernel of dimension 75 $\times$ 2 and $N_A=3$ input and output channels.

Albeit less accurate, this architecture has several advantages. First, it is faster and ${\sim}10\times$ smaller than LAI-Net. Second, when $\lambda_1 > 0$ during training, the output of the first layer, $\hat{y}_1$, represents the probabilities of the linear classifiers. This leads to a more interpretable network since the learned weights of the linear classifiers specify the importance of each SNP to belong at some ancestry.

\section{Experimental Results}
\label{sec:experimental-results}
We used the simulated data previously described to train and test LAI-Net. The network was trained using Adam optimizer and a learning rate of $0.01$ over 100 epochs. The validation set was used to select the training parameters, $\lambda_1$ and $\lambda_2$, and the network hyperparameters: window size, hidden layer size and smoothing kernel size. Table \ref{table:accuracy} presents the accuracy results for chromosome 20 of LAI-Net and Small LAI-Net with and without the smoothing layer compared with the RFMix accuracy.



\begin{table}[h]
  \caption{Accuracy of RFMix, Small LAI-Net and LAI-Net with smoothing (w/ s.) and without smoothing (w/out s.) in the validation and testing set.}
  \label{table:accuracy}
  \centering
  \begin{tabular}{l l c c}
    \toprule
         \multicolumn{2}{l}{\textbf{Method}}     & \textbf{Validation}     &  \textbf{Test} \\
    \midrule
    \multicolumn{2}{l}{\textbf{RFMix \cite{maples2013rfmix}}}       & 92.88\%  & 92.47\%     \\ 
    \multirow{2}{*}{\textbf{Small LAI-Net}} & \textbf{w/out s.}   & 79.70\% & 77.78\%      \\
    & \textbf{w/ s.}                                                & 97.10\% & 96.85\%      \\ 
    \multirow{2}{*}{\textbf{LAI-Net}} & \textbf{w/out s.}         & 82.20\% & 80.29\%      \\
    & \textbf{w/ s.}                                                & 97.96\% & 97.85\%      \\

    \bottomrule
  \end{tabular}
\end{table}

Tests suggest that both LAI-Net and Small LAI-Net are able to achieve state-of-the-art performance. With only two and three layers, the model size of the networks are ${\sim}10$MB and ${\sim}100$MB for Small LAI-Net and LAI-Net respectively. These networks are trained here with data from chromosome 20; their size scales linearly with larger chromosomes. 

\subsection{Missing Data Robustness}
Applications that work with genotype data commonly face data that is noisy or incomplete due to genotyping errors. In other cases only a subset of SNPs might be available due to differing commercial genotyping arrays. Therefore, robustness to missing data is an important element when deploying LAI methods. Current LAI techniques require the user to update the references (training panel) and re-train the model when large numbers of SNPs are missing (eg. when using a genotyping array vs. whole genome sequences or when using different genotyping arrays); our method does not require this.

In order to evaluate the network performance when larger amounts of data are missing, we trained and tested the network with different percentages of missing input SNPs. The structure of the network was not changed and the missing labels were modeled by applying dropout to the input data in both training and testing (i.e. missing SNPs were set to 0). Table \ref{table:missing-data} presents the accuracy values of the estimate on the first and second layer with a different percentage of missing input SNPs.

\begin{table}[htp]
  \caption{Accuracy of LAI-Net for different percentage of missing input SNPs with and without smoothing layer.}
  \label{table:missing-data}
  \centering
  \begin{tabular}{l c c}
    \toprule
        \% \textbf{Missing SNPs}     & \textbf{w/out Smoothing}    &  \textbf{w/ Smoothing} \\

    \midrule
    \textbf{0}      & 80.29\% & 97.85\%     \\ 
    \textbf{25}     & 68.16\% & 95.70\%      \\
    \textbf{50}     & 62.55\% & 94.01\%      \\
    \textbf{75}     & 55.82\% & 92.36\%      \\
    \textbf{90}     & 48.36\% & 87.06\%      \\

    \bottomrule
  \end{tabular}
\end{table}

The accuracy results suggest that the network is able to accurately infer ancestry without a considerable loss of accuracy, even when 50\% of the input SNPs are missing. 
Another advantage is that if only 50\% of the input SNPs are used during deployment, only half of the model parameters need to be stored and only half of the data needs to be processed. This turns missing data from an annoyance into a feature for designing smaller and faster networks that require a fraction of the number of input SNPs as an input.

\subsection{Phasing Errors Robustness}

Humans carry two complete copies of the genome, one from each parent. Current sequencing technologies are typically unable to ascertain whether two neighboring SNP variants belong to the same sequence (maternal or paternal) or opposite sequence. That is, read base-pairs cannot be properly assigned to the paternal or maternal sequences. Assigning variants to their correct sequence is known as phasing, and statistical algorithms have been developed to solve this problem based on observed correlations between neighboring SNP variants allele in reference populations. Such methods include Beagle \cite{browning2007rapid} and SHAPEIT \cite{delaneau2012linear}. However, these tools are not perfect with occasional swaps occurring between the two sequences.

In order to evaluate the network's performance in the presence of phase errors, we trained and tested the network with data containing different percentages of phasing errors. In order to model these errors, we randomly swapped the genomic sequence in locations where the base-pairs differed in the maternal and paternal sequences. In other words, after encoding the SNPs as -1 and 1, the sign of the SNPs in positions where the paternal and maternal are 1 and -1 or vise-versa, were switched with a probability $p$.

Table \ref{table:phasing-data} presents the accuracy results of LAI-Net when different values of $p$ were used for training and evaluation. Results suggest that the network is able to handle small and medium levels of phasing errors, however the accuracy decreases considerably when very high phasing errors ($\sim40\%$) are present.

\begin{table}[htp]
  \caption{Accuracy of LAI-Net with and without smoothing layer for different percentage of phasing error. The networks are trained and evaluated with $p \in \{0, 0.05, 0.1, 0.2, 0.3, 0.4\}$.}
  \label{table:phasing-data}
  \centering
  
  \begin{tabular}{l c c}
    \toprule
        \% \textbf{Phasing Errors}     & \textbf{w/out Smoothing}    &  \textbf{w/ Smoothing} \\
    \midrule
        \textbf{0} & 80.29\% & 97.85\%     \\ 
    \textbf{5}     & 78.61\% & 97.75\%      \\
    \textbf{10}     & 76.94\% & 97.52\%      \\
    \textbf{20}     & 72.98\% & 96.85\%      \\
    \textbf{30}     & 68.18\% & 95.59\%      \\
    \textbf{40}     & 60.64\% & 88.89\%      \\
    \bottomrule
  \end{tabular}
\end{table}

\section{Conclusions}
LAI methods are being used across a broadening array of applications by researchers and practitioners with widely different technical backgrounds. Thus, these methods, besides being accurate, need to be easy to share once trained and must be robust to missing data, allowing for application across differing genotyping platforms. In this work, we present an approach based on neural networks that provides accuracy competitive with state-of-the-art methods and a shareable model that can perform across different genotyping arrays. The ability to share trained models removes the burden of training (and finding appropriate reference populations) from the user, simplifying the use of LAI. Potential pitfalls are reduced, as is the level of experience required of the user, while the training time (the slowest and most computationally expensive step in LAI) need not be born by the user. Most importantly, highly accurate models can be generated on data sets that cannot themselves be shared without breaching privacy restrictions.




\vfill\pagebreak


\bibliographystyle{IEEEbib}
\bibliography{mybiblio}

\end{document}